# Improving Voltage Profile of the Nigerian Power Grid

Amritanshu Pandey *[1], Aayushya Agarwal *[1], Marko Jereminov[1], Tochi Nwachuku[2], Barry G. Rawn[1] and Larry Pileggi[1]

[1]Dept. of Electrical and Computer Engineering
Carnegie Mellon University
Pittsburgh, PA

[2]Klipsch School of Electrical and Computer Engineering
New Mexico State University
Las Cruces, NM

*Abstract-* **Extreme voltages at the system nodes are one of the primary causes for total and partial collapse of the Nigerian grid. In this paper, we develop a framework to re-dispatch the voltage set-points of committed generators in the grid to improve the voltage profile of the system nodes thus lowering the likelihood of a grid collapse. This framework is an extension of circuit-theoretic formulation that can robustly solve general-purpose grid optimization problems. In the results section, we re-dispatch the voltage setpoints for committed generators in real-life Nigerian grid operation and planning test cases to improve the overall voltage profile of the grid. We further demonstrate that the re-dispatched grid is more resilient and secure than the base case through running contingency analyses.**

*Index Terms*—**current voltage formulation, current mismatch formulation, equivalent split-circuit, Nigerian grid, voltage collapse.**

## I. INTRODUCTION

Nigeria is the largest economy in Africa with a GDP of $375.8 billion USD as of 2017. However, the Nigerian grid has underperformed when compared against its peers [1]. The grid has a peak demand forecast of roughly 23960 MW [2]; however, the maximum ever reported peak generation under operation was 5375 MW [2] on the 7th of February 2019. Grid collapses are frequent and its frequency (partial + total) for years 2014 through 2018 is tabulated below[2]:

TABLE 1: FREQUENCY OF TOTAL AND PARTIAL GRID COLLAPSES IN NIGERIA.

| Year | 2014 | 2015 | 2016 | 2017 | 2018 |
|---|---|---|---|---|---|
| Occurrences | 13 | 10 | 28 | 24 | 12 |

Severe grid voltages are one of the many technical issues faced by the Nigerian grid that can cause a grid collapse [3]. Therefore, it can be shown that improving the regulation of grid voltages can reduce the likelihood of voltage collapse. In existing grids in the developed world, operators use a variety of analytical methods to ensure that the grid voltages do not cause a voltage collapse. Generally, the system is dispatched to maintain the grid voltages within a certain pre-defined bound (generally 0.95 p.u. – 1.05 p.u.). In addition, PV-QV curves for the weak nodes are derived to ensure that the grid nodes are not operating close to their critical operating points [4]. Furthermore, transient analysis with more detailed models is often performed to study critical conditions and ensure that a large system disturbance near the weak nodes will not lead to a collapse of the grid [4].

Compared to relatively strong central grids in the US, the existing Nigerian grid is generally considered a weak grid and covers a vast geographical area with relatively few lines. As system load in Nigeria is almost always greater than the generation, the system generally operates in the "load following" mode. In this mode, all operational generation is generally dispatched to meet the maximum possible load without violating system constraints [5]. In doing so, extreme node voltages are often observed that can result in total or partial grid collapse in an event of a large system disturbance.

Improvement in voltage profile of the grid can be achieved through various means amongst which the most common ones include: (i) addition of new equipment such as the reactive power compensating devices in specific locations of the grid through optimal placement algorithms [6] and; (ii) device control through smart strategies with the use of available closed loop control in the grid [7]. However, these methods for improving the voltage profile are generally not viable in the Nigerian grid as the available degree of freedom generally only includes control of generator voltage set-points and manual switching of grid reactors [8].

In this paper, we develop a methodology to improve the voltage profile of the Nigerian grid considering only the restricted available resources. We use circuit-theoretic optimization methods [9]-[10] to constrain the system node voltages within an acceptable range while enforcing generator reactive power limits. Specifically, the developed methodology in the paper re-dispatches the voltage setpoints of the committed generators in the system to operate the voltages at all nodes as close to the predefined band. We use the concept of barrier functions [11] to limit the node voltage variables. To model the barrier function in the circuit-theoretic approach, we use a diode-clamping *like* circuit. The current flowing through these diode-clamping circuits [12] are then penalized in the objective function of the optimization problem to keep the node voltages and generator reactive power within their limits. With this model, there is no penalty if the grid voltages are within a certain band. This corresponds to the diode region where the voltage at the diode terminal is lower than the threshold and thus the diode current is close to zero. In case the node voltages are greater than the pre-defined bounds (given by the threshold), we steeply penalize the objective function thereby attaching a high cost to the node voltages that are operating outside this range, identical to the barrier function. This corresponds to the diode region where the voltage at the diode terminal is able to conduct high diode currents. Using the circuit-based optimization power flow formulation [9] in our prototype tool SUGAR (Simulation with Unified Grid Analyses and Renewables) [13]-[14], we are able to introduce these diode circuits to enforce the bus voltage and generator reactive power limits by minimizing the diode currents.

In Section II of this paper, we discuss the previously proposed circuit theoretic framework for steady-state analysis of power grid [13] including adjoint circuits [9] that are used for general optimization analyses of the grid. Section III extends the existing framework with the inclusion of mathematical constraints that model the optimization circuits to enforce the limits for grid node voltages and generator reactive power limits. Section IV will describe the proposed approach along with implementation within in our tool SUGAR to ensure convergence for these models and the final section discusses the results for an operational Nigerian grid test case with roughly 5500 MW load dispatch and a planning Nigerian grid test case with roughly 7000 MW load dispatch.

## II. BACKGROUND

### A. Current-Voltage Approach to Power Flow Analysis

Current-voltage (I-V) based formulations have been explored in the past for performing power flow and three-phase power flow analyses [13]-[16]. Amongst these approaches, a recently introduced equivalent circuit approach [13] maps the different network models of the grid (e.g. PV, PQ, etc.) into their respective equivalent circuits and further aggregates them together to create the whole network model of the grid to solve for the node voltages and branch currents. This circuit-based formulation represents both the transmission and distribution power grid model as an aggregation of circuit elements. In following this approach, this method preserves the physical nature of the grid elements to result in convergence to the correct physical solution, while improving robustness through the use of circuit simulation techniques.

One of the circuit simulation methods limits the Newton-Raphson (NR) step to avoid leading the solution to a space that may cause divergence or oscillations. In [13], we demonstrated that damping the step size of each element based on the physical properties results in better convergence of the solver. While limiting the NR step can improve convergence, limiting by itself does not guarantee that the solution represents a physically stable dispatch. In order to lead the NR to a meaningful solution (high voltage magnitude and angular stable), we introduced a homotopy method [13], *"Tx stepping"*, that sequentially solves a series of sub-problems with the use of a homotopy factor. The Tx-stepping method first solves a trivial problem and uses the solution to iteratively solve a series of sub-problems that eventually is equivalent to the original problem. This approach has demonstrated robust convergence to physically meaningful solutions.

### B. Optimization Using Adjoint Circuits

The equivalent circuit approach for steady-state analysis of the grid can be extended to formulate equality constrained optimization problems via the use of adjoint theory [9]. In this approach, in addition to the power flow circuits, adjoint circuits are added to the framework to represent the necessary first-order optimality conditions for the optimization problem.

A general form of equality constrained grid optimization problem for power system analysis is as follows:

$$\min_{x} \quad F(\boldsymbol{x}) \qquad \qquad (1)$$
$$\text{subject to} \quad h(X) = 0$$

where $h(X)$ represents the non-linear and linear system network constraints and $X$ represents the vector voltages and additional state variables in the system. $\boldsymbol{x} \subseteq X$ is a vector of variables in whose solution space the objective function is minimized.

In order to solve the problem given by (1) using the equivalent circuit approach [9], the necessary first-order optimality conditions given by the KKT conditions are mapped onto a set of equivalent circuits. These circuits are represented by a set of power flow circuits that can be derived following the methodology in [13] and a set of adjoint circuits that can be derived following the methodology in [9]. The solution of this net aggregated circuit (power flow circuits + adjoint circuits) then corresponds to an optimal solution of the optimization problem assuming the hold of the second-order sufficient conditions and is obtained by using the circuit simulation techniques described in [9],[17].

*1) Inequality Constraints*

The circuit-based approach for equality constrained optimization problems can also be extended to solve power grid optimization problems with inequality constraints [9]. In such a constrained optimization problem, in addition to ensuring that the network equations are met, we constrain certain variables within bounds, $d$ using a complementary slack variable $s$, defined by:

$$g(V) + s = d \qquad (2)$$
$$s \geq 0 \qquad (3)$$

A common technique to solve for (2) and (3) in the optimization solvers is to use barrier methods [9].

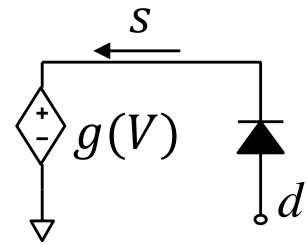

*Figure 1: Diode approximation of complimentary slack conditions.*

Barrier methods utilize functions that constrain the slack variable to be within the bounds by making the system infeasible if the slack violates the limits. We can use an ideal diode model to approximate the behavior of the barrier function, as shown in Figure 2 with a lower bound of 0.9 pu.

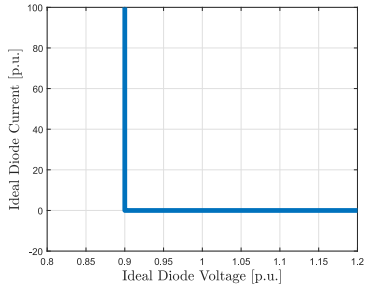

*Figure 2: Voltage-current characteristic of an ideal diode model.*

Typically, the current flowing through an ideal diode is represented by a piecewise function that conducts an infinite amount of current if the voltage variable exceeds the threshold voltage [12] as shown in Figure 2, and behaves similarly to the slack variable in the barrier function.

### III. VOLTAGE PROFILE CONSTRAINING FUNCTIONS

In our framework for analyses of the Nigerian grid, we introduce the slack variables to incentivize the grid voltages for being in a pre-defined range while ensuring that the machine reactive power limits are not violated and the optimal solution is feasible [9], which can be formulated as:

$$V_{MIN} \leq V_{SET} \leq V_{MAX} \quad (4)$$
$$Q_{MIN} \leq Q \leq Q_{MAX} \quad (5)$$

This is used to constrain the system node voltages ($V_{SET}$) between $V_{MAX}$ and $V_{MIN}$ and the reactive power of the generator ($Q$) between $Q_{MAX}$ and $Q_{MIN}$.

In the circuit-theoretic approach for optimization, we apply the approximations of the behavior of this barrier function with ideal diodes circuit as shown in Figure 3. This circuit is analogous to adding diode voltage clamping circuits in analog circuits to limit the voltages.

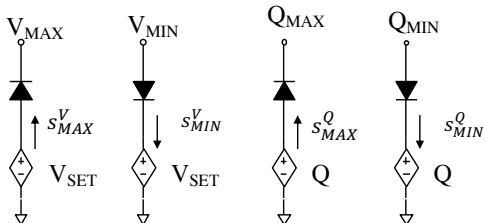

*Figure 3 Diode limiting circuits for limiting voltages and reactive powers between bounds.*

The current flowing through the controlled source, $s^V_{MAX}$ or $s^V_{MIN}$ represents the mismatch between the voltage at a node, $V_{SET}$ and the maximum and minimum bounds respectively. Similarly, $s^Q_{MAX}$ or $s^Q_{MIN}$ represents the mismatch between the reactive power, $Q$ and its respective bounds. Minimizing this current in the optimization framework mimics the barrier method. When the controlled variable is within the maximum and minimum bounds (represented by $V_{MAX}$ or $Q_{MAX}$ and $V_{MIN}$ or $Q_{MIN}$ respectively), the current flowing through the diode circuit is approximately zero. However, if the controlled voltage is outside the bounds, the current through one of the diodes will conduct depending on whether the upper bound or lower bound has been violated.

The current flowing through an ideal diode is represented by a piecewise equation that represents hard limit bounds on the controlled variable in our formulation and has difficult convergence properties as Newton-Raphson (NR) cannot be directly applied to such functions. In our problem methodology, we implement both hard limits on problem variables as well as soft limits. The hard limits are useful to enforce physical limits such as the reactive power output of the generator. In contrast, soft-limits are useful for enforcing desirable behavior such as better grid voltages. Soft-limits are enforced by formulating the diode current such that it has a finite value when outside the range. When soft-limits are applied towards improving bus voltage magnitudes, they add zero cost to the objective function (i.e. zero diode current) when they are in the desired range and penalized when they are outside (i.e. high diode current). These hard and soft limit behaviors are represented below through derivations of models that can be directly used with NR algorithms.

*1) Exponential Diode Current Model*

To overcome the challenges faced by the NR algorithm for the piecewise model of diodes, circuit simulation models for the same using exponential functions are used, as shown in red in Figure 4. In our formulation, the exponential model is given by (6) and it penalizes the operation of the model parameter outside the defined threshold limits. In the exponential model, $I_o$ represents the weight associated with the optimization penalty, and $\kappa$ dictates the steepness of the function.

$$s^V \equiv s^V_{MIN} - s^V_{MAX} = I_o\left(e^{\kappa(V_{MIN}-V_{SET})} - e^{\kappa(V_{SET}-V_{MAX})}\right) \quad (6)$$

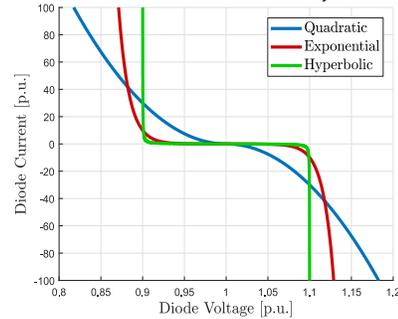

*Figure 4: Diode current model approximations.*

*2) Hyperbolic Diode Current Model*

While exponential models are used in circuit simulation, they are often difficult to converge due to numerical overflow. As a result, to represent the hard limits on generator reactive powers, we model the diode currents using a small $\epsilon$ (of 1e-4) in the hyperbolic function (7) as shown in Figure 4 (green). This function was first presented in [10] for constrained optimization problems. Furthermore, the hyperbolic function is differentiable and is preferred in our formulation for representing the hard limits.

$$s^Q \equiv s^Q_{MIN} - s^Q_{MAX} = \frac{\epsilon}{Q - Q_{MIN}} - \frac{\epsilon}{Q - Q_{MAX}} \quad (7)$$

*3) Quadratic Diode Current Model*

To enforce soft-limit constraints in our framework, we use quadratic diode current model. These constraints allow operation of the model parameters outside its range yet penalizes them for doing so. A quadratic function given in (8) models this behavior. Figure 4 (blue) shows the diode current characteristics as a function of diode voltage for this soft limit model. The steepness of the quadratic function is determined.

by $a$ which controls the weight of the penalty in the objective function for operating outside of the bounds.

$$s^V \equiv s^V_{MIN} - s^V_{MAX}$$
$$= \begin{cases} a\left(V_{SET} - \frac{V_{MAX}-V_{MIN}}{2}\right)^2, & V_{SET} < \frac{V_{MAX}-V_{MIN}}{2} \\ -a\left(V_{SET} - \frac{V_{MAX}-V_{MIN}}{2}\right)^2, & V_{SET} \geq \frac{V_{MAX}-V_{MIN}}{2} \end{cases} \quad (8)$$

## IV. PROBLEM FORMULATION

The objective of this paper is to improve the voltage profile of the grid nodes. In our approach, this objective is met via minimizing the currents $\boldsymbol{s} = \{s^V_{MAX}, s^V_{MIN}, s^Q_{MAX}, s^Q_{MIN}\}$ of the limiting circuits developed in Section III of this paper such that the network constraints given by Kirchhoff's current laws are satisfied. Mathematically, the optimization problem can be written as follows:

$$\min_X \quad \|\boldsymbol{s}^Q_{MAX}, \boldsymbol{s}^Q_{MIN}\|_2^2 + \|\boldsymbol{s}^V_{MAX}, \boldsymbol{s}^V_{MIN}\|_2^2$$
$$\text{s.t.} \quad h(V) = 0$$
$$F_{SL}(V_{SET}, V_{MAX}, V_{MIN}, s^V_{MAX}, s^V_{MIN}) = 0 \quad (9)$$
$$F_{HL}(Q_{SET}, Q_{MAX}, Q_{MIN}, s^Q_{MAX}, s^Q_{MIN}) = 0$$
$$(V_{SET})^2 - (V_R)^2 - (V_I)^2 = 0 \equiv V_{EQ}$$

The feasible region for the grid voltages ($V_{SET}$) is given by the vector of $V_{MAX}$ and $V_{MIN}$ whereas limits for the set of generator's reactive power ($Q$) is given by the vector of $Q_{MAX}$ and $Q_{MIN}$. The non-linear equations for soft-limit bus constraints and hard-limit generator constraints are given by $F_{SL}$ and $F_{HL}$, respectively and are described in Section III of this paper. Similar to in the case of network models (like PV and PQ) in [9] and [13], these equations are linearized to represent the corresponding primal and dual circuits that are further used in Newton's method below.

This problem as described above is a non-convex quadratic programming problem and is solved in our approach using Newton's method [9] to obtain a local optimum solution. In order to avoid the "non-useful solutions" due to non-convexity of the problem, we make use of limiting and homotopy methods [13] to ensure that the local optimum solution is obtained such that physical nature of the grid is maintained while pushing the solution towards the high voltage solution. To solve the problem using Newton's method, we solve for the set of equations described by the necessary first-order optimality conditions. To derive that we first derive the Lagrangian of the problem given by:

$$\mathcal{L}(V_{TOT}, \lambda_{TOT}) = \left(\|\boldsymbol{s}^Q_{MAX}, \boldsymbol{s}^Q_{MIN}\|_2^2 + \|\boldsymbol{s}^V_{MAX}, \boldsymbol{s}^V_{MIN}\|_2^2 + \lambda^T(h(V)) + \lambda^T_{SL}F_{SL} + \lambda^T_{HL}F_{HL} + \lambda^T_{VEq}V_{EQ}\right) \quad (10)$$

In the equations below, we define $V_{TOT}$ as the set of unknown variables that we are solving for:

$$V_{TOT} = \{Q, V, V_{SET}, s\} \quad (11)$$

and the Lagrangian multipliers as:

$$\lambda_{TOT} = \{\lambda^T, \lambda^T_{SL}, \lambda^T_{HL}, \lambda^T_{VEq}\} \quad (12)$$

The necessary first-order optimality conditions can be derived from the Lagrangian function and are given by KKT conditions:

$$\frac{\partial \mathcal{L}}{\partial V_{TOT}} = 0 \quad (13)$$
$$\frac{\partial \mathcal{L}}{\partial \lambda_{TOT}} = 0 \quad (14)$$

To solve for the set of non-linear equations defined by the KKT conditions, we first linearize the equations (13) and (14). In our approach, the linearized terms are then mapped into their respective equivalent circuits, using techniques introduced in Section II. Primal circuits are used to map the terms in (14) while dual circuits are used to map the terms in (13). Together, the combined primal and dual circuits completely map the linearized terms in (13) and (14) that further represent the necessary first-order optimality conditions for the equality constrained optimization problem. Importantly these terms are updated in each step of Newton's method to solve for the non-linear equations.

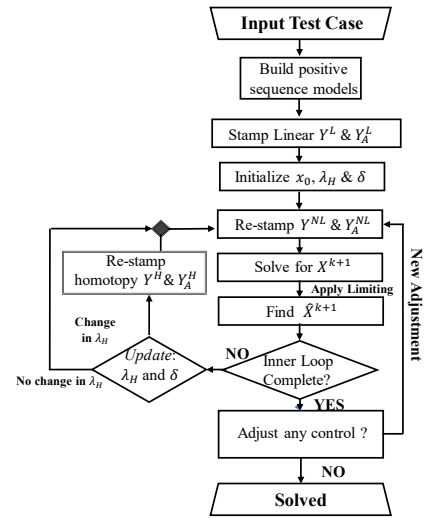

*Figure 5: Flowchart for implementation of voltage profile improvement algorithm.*

Figure 5 shows the recipe for solving the formulated optimization problem following the approach in [9] and [13] that is further extended with the use of novel models developed in Section III of this paper. This framework makes use of circuit simulation methods [9], [13] to ensure robust convergence of the proposed methodology, which otherwise would have been hard to achieve with such highly non-linear models. The solver starts with building the system primal and dual models based on the input file supplied. Both primal and dual (adjoint) linear models ($Y^L, Y^L_A, J^L$) are then stamped (added to the system matrix) in the Jacobian matrix. Input state variables and other solver parameters ($x_0, \delta, \zeta, \lambda$) are then initialized. Non-linear primal and dual models are then stamped iteratively ($Y^{NL}, Y^{NL}_A, J^{NL}, J^{NL}_A$) and NR is applied with limiting methods to calculate the next iterate for unknown state variables ($\hat{X}^{k+1}$). Homotopy and limiting parameters are then dynamically updated and homotopy models ($Y^H, Y^H_A, J^H, J^H_A$) are stamped or re-stamped if required to ensure convergence.

## V. RESULTS

In this section, we demonstrate that the problem formulation described within this paper can be used to improve the voltage profile of the Nigerian grid and therefore can reduce the likelihood of voltage collapse due to either systemic low voltages or high voltages in the grid. We use our prototype tool SUGAR to run the analysis following the methodology described in Section IV of the paper and optimize for the re-dispatch of generator voltage set-points to have the bus voltages between 0.85 pu and 1.15 pu [8] while enforcing physical reactive power limits of the generators. Using the diode models presented earlier, we used soft-limit constraints for improving the profile of the node voltages and hard-limit constraints for enforcing the generator reactive power limits. In this section, we use two real-life test cases for the Nigerian grid: (i) an operations model serving 5500 MW and (ii) a model including future reinforcements that may allow serving up to 7000 MW. As the operational Nigerian test case represents a well-behaved system, we use it for validation of our methodology. The reinforced 7000 MW was initially observed to have high voltage issues, which are shown in this paper to be significantly relieved when re-dispatched based on our methodology of setpoint optimization.

### A. 5000 MW Operations Case

The 5000 MW is a well-behaved operational test-case and therefore serves as a good example for validating our approach. For this network, we expect a re-dispatch based on our methodology that is similar to the base dispatch. This is shown in Figure 6. As can be seen from the figure, the grid voltages for the base case and re-dispatch case are similar thereby validating our approach for dispatching of resources.

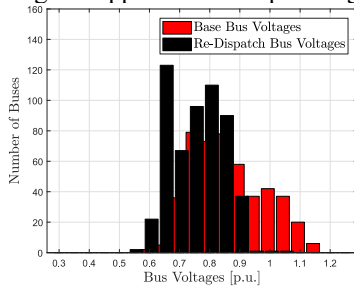

*Figure 6: Grid voltages for the base and re-dispatched 5000 MW testcase.*

Although, it is noteworthy that some of the voltages for the re-dispatch case are seen to be lower than that of the base case in Figure 6; however, as it can be seen from the contingency results in Figure 7 that on an average the re-dispatch contingency instances have a higher minimum voltage (min. across all nodes in that instance) than that of the base case thereby essentially showing that overall the re-dispatch is more voltage-secure than that of the base case.

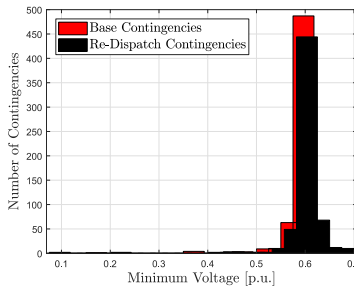

*Figure 7: Minimum voltages for contingency instances for base and re-dispatch test cases.*

### B. 7000 MW Planning Case

#### 1) Base case analysis

The reinforced 7000 MW planning case represents a very high system loading network and contains high voltages that are practically un-dispatchable. The **red distribution** in Figure 8 plots the node voltages in the base dispatch for this network file. It is observed that the base case has voltages as high as 1.9 pu which makes it un-dispatchable as protective relaying would trip under such high voltages that would most likely result in a partial or total collapse of the grid. Using our optimization methodology, we were able to re-dispatch the generator voltage setpoints in the grid and improve the voltage profile of the case as shown by the **black distribution** in Figure 8. It is apparent that the voltages in the re-dispatched case are improved and have max and min voltages in the range of 0.7 pu to 1.3 pu, thereby significantly decreasing the likelihood of voltage collapse due to the tripping of protective relaying or other technical causes.

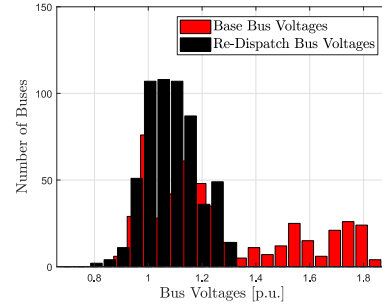

*Figure 8: Base and Re-dispatched case bus voltages.*

In addition, the figure below shows the generator voltage setpoints for the base and re-dispatched test cases.

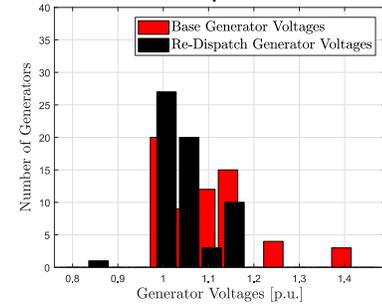

*Figure 9: Generator voltage set-points for base versus re-dispatched testcase.*

#### 2) Contingency Analysis

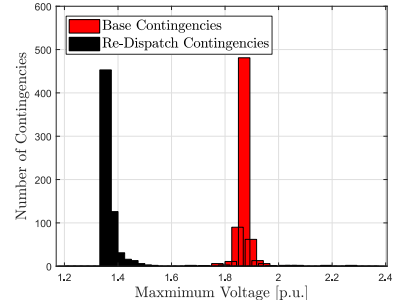

*Figure 10: Distribution of maximum voltages observed during N-1 contingency analysis of the 7000 MW case.*

The base 7000 MW case suffers from high voltages at system nodes and hence the majority of contingencies also have high voltage levels at which the grid cannot operate. This is shown in Figure 10 wherein the distribution of maximum bus voltage that is observed for each N-1 contingency case is plotted. It is shown that the contingencies of the base case (shown in **red**) are un-dispatchable with maximum voltages around 1.9 pu and would almost certainly result in the protective relaying trip. However, when the case is re-dispatched using the optimization framework, the distribution of maximum voltages (shown in **black**) across multiple contingencies is significantly improved thereby decreasing the likelihood of total or partial voltage collapse of the system if the case is dispatched as-is. This shows that the optimization framework was not only able to improve the voltage profile but also decrease the risk of collapse during a contingency.

*C. 5950MW Planning Case*

While the optimization formulation was able to redispatch the planned 7000MW planning case with a better voltage profile, it may not still meet operational guidelines of the Nigerian grid. The real 330 kV node voltages, observed in the daily National Operating Reports (NOR), demonstrate that the 330kV buses typically lie below 1.15 pu and above 0.75 pu [8] (Figure 11). For the investments represented by the planning case to be operational today, the voltages of the 330kV buses must also lie within the operational band.

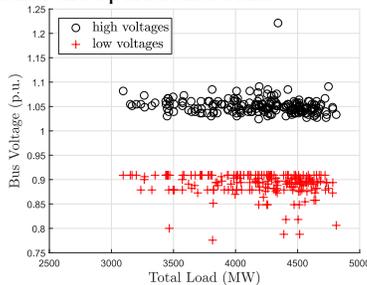

*Figure 11: High and Low voltages reported from NOR Reports [8].*

To make the 7000MW case operational, the robust optimization formulation was able to scale the load uniformly to satisfy the additional voltage constraints on the 330kV buses. After scaling the load uniformly by 85% (making it a 5950MW planning case), the voltages of the 330 kV buses were shown to be in the range of voltages as observed in the operating Nigerian grid, as shown in Figure 12.

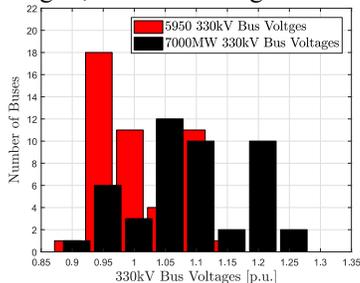

*Figure 12: 330kV bus voltages of 5950MW and re-dispatched 7000MW cases.*

## VI. CONCLUSIONS

The Nigerian grid is known [8] to experience some extreme voltages at some nodes, which indicate the underlying grid stress that contributes to a high volume of partial or total grid collapses in recent years. In this paper, we address the voltage profile concerns present at transmission level by proposing a formulation for re-dispatching of limited grid resources based on the optimization framework using the equivalent circuit approach. The results shown indicate that the robust optimization formulation can be used as a planning tool to support serving higher loads, while carefully considering the voltage profile.

## VII. ACKNOWLEDGMENT

This work was supported in part by the Defense Advanced Research Projects Agency (DARPA) under award no. FA8750-17-1-0059 for RADICS program, and the National Science Foundation (NSF) under contract no. ECCS-1800812.